\documentclass[epj]{svjour}
\usepackage{graphics}
\usepackage{graphicx}
\usepackage{dcolumn}
\usepackage{bm}
\usepackage{amsmath}
\usepackage{subfig} 	   
\usepackage{hyperref}  
\usepackage{cite}         

\begin{document} 
\title{Mitigation of Electron Cloud Effects in the  FCC-\lowercase{ee} Collider} 

\author{Fatih Yaman\inst{1}\thanks{\emph{Corresponding author:} fatihyaman@iyte.edu.tr \\ \textbf{Preprint submitted to EPJ Techniques and Instrumentation}} \and Giovanni Iadarola\inst{2} \and Roberto Kersevan\inst{2} \and Salim Ogur\inst{3} \and Kazuhito Ohmi\inst{4} \and Frank Zimmermann\inst{2}  \and Mikhail Zobov\inst{5}  
}                     
\institute{Izmir Institute of Technology, Izmir 35430, Turkey 
\and CERN, 1211 Geneva 23, Switzerland
\and CNRS/IJCLab, 91400 Orsay, France 
\and KEK, Tsukuba, Ibaraki 305-0801, Japan
\and INFN LNF, via Enrico Fermi 40, 00044 Frascati, RM, Italy
}
\date{Received: date / Revised version: date}

\abstract{Electron clouds forming inside the beam vacuum chamber due to photoemission and secondary emission may limit the accelerator performance. Specifically, the electron clouds  can blow up the vertical emittance of a positron beam, through a head-tail-type single-bunch instability,  if the central electron density exceeds a certain threshold value, that can be estimated analytically. Using the codes PyECLOUD and VSim, we carried out detailed simulations of the electron-cloud build up for the main arcs and the damping ring of the FCC-ee collider, in order to  identify the effective photoemission rate and secondary emission yield required for achieving and 
maintaining the design emittance.  
To this end, we present the simulated electron density at the centre of the beam pipe for various bunch spacings, secondary emission yields, and photoemission parameters,
in the damping ring and in the arcs of the collider positron ring. 
To gain further insight into the underlying dynamics, the obtained spatial and energy distributions of the cloud  electrons  are illustrated as a function of time. In addition, we compare results obtained for two different secondary emission models (``Furman-Pivi'' and ``ECLOUD''), thereby indicating the uncertainty inherent in this type of study, without any prototype vacuum chambers yet 
available.    
We also point out a few situations where the two 
secondary-emission models yield similar density values.
Finally, based on our simulation results for two different design variants, we conclude that the new parameter baseline
of the FCC-ee will facilitate electron-cloud mitigation. 
} 
\maketitle

\section{Introduction}
The future high-energy circular 
electron-positron collider FCC-ee 
is the first stage of the integrated  
FCC project proposed at CERN  
\cite{Benedikt_et_al,Mangano_et_al}, and based on a new 
$\sim100$ km tunnel infrastructure.
The FCC-ee shall investigate  
open questions in modern particle physics by operating 
at several different collision energies between $88$ and $365$~GeV 
In addition to the double-ring collider, sharing the tunnel with a full-energy booster synchrotron serving for top-up injection, 
the FCC-ee requires a pre-injector complex.
This pre-injector complex consists of a linac, damping ring (DR), pre-booster and top-up booster \cite{Ogur_et_al,Ogur_thesis}. 
The function of the DR is to reduce the emittance of the positron beam at an energy of $1.54$~GeV in a sufficiently short time scale.

In the vacuum beam pipes of accelerators and storage rings,
the primary electrons are generated by photoemission due to synchrotron radiation, by the ionization of the residual gas, and, possibly, also by   
the uncontrolled loss of stray beam particles.  
In addition, and importantly, further, 
``secondary'' electrons are produced 
when primary electrons of sufficient energy 
hit the pipe wall \cite{Furman_Lambertson,Zimmermann_Ecloud_code}, which can lead to an amplification. 
The possibly resulting exponential generation of electrons may cause both incoherent emittance growth and coherent beam instabilities.

In the early 1990s, a pioneering simulation of electron cloud build up, due to photoemission, could explain observations of coupled-bunch beam instabilities in positron operation 
at the KEK Photon Factory \cite{Ohmi_95}.
A few years later,  
the secondary emission process was added in early electron-cloud simulations for the designs of the PEP-II B factory \cite{lambertson} 
and the Large Hadron Collider (LHC) \cite{fzlhc}.  
Around the year 2000,
models of a single-bunch head-tail instability driven 
by an electron cloud were developed 
to explain the observed vertical blow up of  
the positron beam in the KEKB B factory collider \cite{kofzprl,kofzeppre}.  
A recent comprehensive review article  \cite{Shiltsev_and_Zimmermann} presents an overview of the past, present and future charged particle colliders, in almost all of which 
electron clouds might occur. 
A historical perspective of 
modeling and simulation efforts for the electron-cloud build-up mechanism, 
along with diagnostics and mitigation of the resultant instability, can be found in Ref.~\cite{Zimmermann_18}.

Our present work aims at analysing electron cloud build-up scenarios for the FCC-ee DR and for the FCC-ee collider arcs, considering two different models for the secondary 
emission yield, and scanning surface parameters in order to
identify maximum acceptable values for primary photoelectron rate and maximum secondary emission yield. 
We also compare the impact of recent changes to the FCC-ee baseline parameters on the likelihood of electron-cloud formation. 
Similar comparative studies were carried out in the past for proton beams in the LHC, e.g.~by G.~Bellodi\cite{Bellodi}
using the two codes  POSINST~\cite{Furman_Lambertson} 
and ECLOUD~\cite{Zimmermann_Ecloud_code}. 

\section{Simulation}
In this article, we employ  the two codes PyECLOUD~\cite{PyECLOUD} and VSim~\cite{VSim} to perform 2D electrostatic particle in cell (PIC) simulations
of the electron-cloud build up process.
The computational domain models a circular, copper vacuum chamber with the pipe radius $35$~mm for the collider arc dipole
while the beam-pipe radius for the DR is varied over 
the range of $10$--$30$~mm.
Furthermore, $0.01415$~T and $1.8$~T external magnetic fields are included along the transverse direction in the beam pipe
of collider and DR, respectively.
The 142 G field for the collider arc magnets corresponds to operation on the Z pole (45.6 GeV per beam), where the beam current is highest,
and which, due to the short bunch spacing, 
should be most susceptible to electron-cloud formation. 
For the DR, the 1.8 T field represents the strength of the damping wigglers, while the DR 
arc dipoles have a lower (but still high) 
field of 0.66 T \cite{Ogur_thesis}. 
The beam sizes of beam injected into the DR widely differ
from those at extraction. Therefore, we examine either case. 

Two different models for the secondary emission yield of copper 
are used for our study.  
The Furman-Pivi secondary electron yield model 
is based on measurements for PEP-II 
in California \cite{FPmodel,Wullf_Iadarola}, 
while the ECLOUD model was constructed from 
laboratory measurements at CERN for the copper surface of the 
LHC~\cite{Hilleret,Iadarola_Thesis}.

\begin{table}[htbp]
\centering
\caption{Parameters employed for the simulations 
of the damping ring \protect\cite{Ogur_thesis} and
collider arcs for the CDR parameters with 2 Interaction Points (IPs) \protect\cite{Benedikt_et_al}
and the new baseline with 4 IPs \cite{FCCISbase}. 
Collider bunch length is shown as due to synchrotron radiation and including the effect of beamstrahlung 
(``BS'', in parentheses).  
\label{Parameters}}
\begin{tabular}{l c c c c}
\hline
\\ [-0.15cm]
                                           &\hspace{0.5cm} DR               \hspace{0.5cm}    &DR               \hspace{0.5cm}             &
                                           Collider Arc 	     Dipole              \hspace{0.5cm}         &Collider Arc \\ [-0.1cm]
                                           &\hspace{0.5cm} Injection       \hspace{0.5cm}   &Extraction      \hspace{0.5cm}             & (CDR, 2 IPs)                  \hspace{0.5cm}         & Dipole (4 IPs) \\[0.05cm]  
\hline
\\ [-0.2cm]
beam energy [GeV]       		    \hspace{0.1cm}    	&1.54                            &\hspace{-0.8cm} 1.54                                &\hspace{-0.8cm}45.6        		       &45.6	         	         \\[0.1cm] 
bunches per train  	    		    \hspace{0.1cm}   	&2                                 &\hspace{-0.8cm} 2                                     &\hspace{-0.8cm}150          		       &150			         \\[0.1cm] 
trains per beam              	            \hspace{0.1cm}  	&8                                 &\hspace{-0.8cm} 8                                     & \hspace{-0.8cm}1           		        &1			         \\[0.1cm] 
r.m.s. bunch length [mm]              \hspace{0.1cm}  	&3.4                              &\hspace{-0.8cm} 2.1 &\hspace{-0.8cm}3.5         		        &4.32 		          \\[0.1cm] 
              \hspace{0.1cm}  	&                              &\hspace{-0.8cm}  &\hspace{-0.8cm} (12.1 w.~BS)       		        & (15.2 w.~BS)		          \\[0.1cm] 
hor.~r.m.s. beam size [$\mu m$]      \hspace{0.1cm}      &2200                           &\hspace{-0.8cm} 98                                   &\hspace{-0.8cm}120        		        &207			          \\[0.1cm] 
vert.~r.m.s. beam size [$\mu m$]       \hspace{0.1cm}     &2800                           &\hspace{-0.8cm} 47                                   &\hspace{-0.8cm}7              		        &12.1		           \\[0.1cm] 
external magnetic field [T]            \hspace{0.1cm}     &1.8                              &\hspace{-0.8cm} 1.8                                   &\hspace{-0.8cm}0.01415    		         &0.01415	                    \\[0.1cm] 
bunch population $N_b$  [$10^{11}$]  \hspace{0.1cm}                             &$ 0.22$     &\hspace{-0.8cm} $0.22$        &\hspace{-0.8cm}$1.7$    &$2.76$    
\\ [0.1cm]
circumference $C$ [km]  \hspace{0.1cm}    &  0.242  & \hspace{-0.8cm}  0.242  & \hspace{-0.8cm} 97.76  &  91.2 \\[0.1cm] 
momentum compaction factor $\alpha_{C}$ [$10^{-4}$]
 \hspace{0.1cm}    & 15 & \hspace{-0.8cm}  15 & \hspace{-0.8cm}  0.148  & 0.285\\[0.1cm] 
synchrotron tune $Q_s$  \hspace{0.1cm}     & 0.022 & 
\hspace{-0.8cm} 0.022 & \hspace{-0.8cm}  0.025 & 0.037 \\[0.1 cm]
average beta function $\beta_y$ [m]  \hspace{0.1cm}     & 1.5 & 
\hspace{-0.8cm} 1.5 & \hspace{-0.8cm}  50 & 50 
\\[0.1 cm]
threshold density $\rho_e$ [$10^{12}$~m$^{-3}$]  \hspace{0.1cm}     & 966.03 & 
\hspace{-0.8cm} 27.11 & \hspace{-0.8cm}  0.027 & 0.043
\\
\\ [-0.2cm]
\hline
\end{tabular}
\end{table}

References~\cite{Ohmi2016, Belli_et_al_2018, Ohmi2019} described earlier electron-cloud studies of the FCC-ee,  
assuming older parameters and applying different approaches. 
Here, we consider the machine and beam parameters listed in Table~\ref{Parameters}.

The threshold central electron density 
for the single-bunch instability can be estimated 
as \cite{Ohmi2016,pre}
\begin{equation}
    \rho_{\rm thr} = \frac{2 \gamma Q_{s} \omega_e \sigma_z/c}{\sqrt{3} K Q r_e \beta_y C} \; ,  
    \label{rhothr} 
\end{equation}
where 
\begin{equation}
\omega_{e}= \left( \frac{N_{b} r_e c^2 }{\sqrt{2 \pi} \sigma_z \sigma_y (\sigma_x +\sigma_y)} \right)^{1/2}\; ,
\label{omegae}
\end{equation} 
$K=\omega_e \sigma_z/c$,
 $Q= {\rm min} (\omega_e \sigma_z/c, 7)$.
The estimated threshold values according to (\ref{rhothr}), for colliding beams  (i.e., with beamstrahlung), 
are also included in Table \ref{Parameters} (bottom row).

For the FCC-ee damping ring, a definite value for the 
beam-pipe radius
has not yet been chosen. This parameter 
could be optimized in view of electron cloud.  Therefore, 
we consider beam-pipe radii of $10$, $20$, and $30$~mm, a  
total secondary emission parameter between $1.1$ and $2.1$, 
and initial electron densities of 
$10^{10}$, $10^{11}$, or $10^{12}$~m$^{-3}$.
On the other hand, 
for the FCC-ee collider, the beam-pipe radius has been
fixed at 35 mm (with additional narrow 
horizontal winglets to remove and absorb most of the synchrotron-radiation photons).   
In simulations for the FCC-ee collider arcs, 
we vary the bunch spacing, 
the total SEY values and the photoemission rates,
and compare results for the
two different secondary emission yield models.

In the simulation, 
cloud electrons are represented by macro particles. 
At each time step we solve Poisson's equation  on a uniform two-dimensional Cartesian grid. 
The accuracy and the convergence of the solution are evaluated by scanning the number of macroparticles and 
the size of the grid cells~\cite{Yaman_talk}. 
PyECLOUD employs an adaptive scheme to control the number of electrons per macro particle during the simulation  \cite{Iadarola_Thesis}.
We choose the number of physical particles per macro particle to be on the order of $10^5$. The macroparticles are 
distributed on a square grid.  
Thereby, we follow the approach adopted in the study 
\cite{SethIPAC05}, which investigated the electron cloud build-up for the PIP-II using VSim simulations.

The number of primary electrons generated by a single positively-charged particle per 
unit length, $n_\gamma^\prime$, is 
\begin{equation}
n_\gamma^\prime = Y_\gamma \, \frac{5 \, \alpha \, \gamma}{2 \sqrt{3}\rho} \, , 
\end{equation}
where $\alpha \approx 1/137$ denotes 
the fine-structure constant, $\gamma$  the Lorentz factor
($\gamma \approx 10^5$ for the collider on the Z pole,
and $\gamma\approx 3000$ for the DR), and 
$\rho $
the radius of curvature of the particle path~
($\rho \approx 11000$~m for the collider arc dipoles,
$\rho \approx 3$~m 
for the DR dipole) 
\cite{Zimmermann_Ecloud_code}. 
The photoelectron yield coefficient $Y_\gamma  $ is traditionally 
considered to be about $0.1$, i.e.~0.1 photoelectrons emitted per absorbed photon.
However, for the FCC-ee collider arcs the antechamber will remove a large fraction 
of the synchrotron-radiation photons from the beam-pipe proper, which will 
lead to a reduction of the effective value for the yield $Y_\gamma  $.
Therefore, in this study, we scan the photoemission rate $n_\gamma^\prime$ 
from $10^{-3}$~m$^{-1}$ down to $10^{-6}$~m$^{-1}$ for the collider arc dipoles. 

One of the main ingredients for the electron cloud build-up 
simulations is the secondary emission model.
For the first model, the so-called Furman-Pivi model, 
the SEY parameters consist of three components 
taking into account, respectively, 
the contributions of the elastically backscattered, rediffused and true-secondary electrons~\cite{FPmodel}. 
The Furman-Pivi model is available both in PyECLOUD and VSim. 
Preliminary comparisons of results obtained from these two codes 
were presented recently \cite{Yaman_et_al_FCCWeek2021,Yaman_et_al_FCCWorshop2021}.

\begin{figure}[!tbh]
    \centering
    \includegraphics*[width=0.6\textwidth]{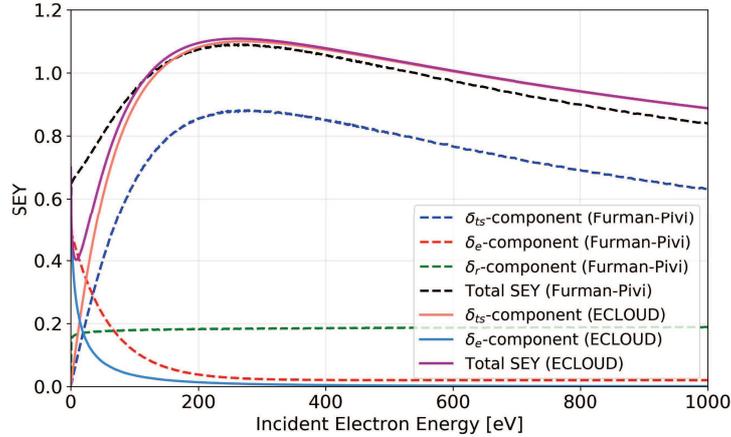}
		\caption{Composition and overall shape for the two SEY models. }
   \label{fig:SEY_Components}
\end{figure}

In our numerical study,
we adjust the value of true-secondary component $\hat{\delta}_{ts}$ 
to obtain a particular total SEY value, namely 
\begin{equation}
\hat{\delta}_{t} \simeq \hat{\delta}_{ts} + 0.22 \quad \mbox{provided that} \quad \hat{E}_{ts} \, \gg \, \hat{E}_e, \, E_r \, , 
\end{equation}
with $ \hat{E}_{t} \simeq \hat{E}_{ts} = 276.8$~eV, $\hat{E}_e = 0$~eV, $E_r = 0.041$~eV, 
for the PEP-II surface pipe sample \cite{FPmodel, Yaman_talk}. 
The variation of SEY components 
with respect to incident electron energy 
for SEY=1.1 is illustrated in Fig.~\ref{fig:SEY_Components}.
Even if the total secondary electron yield parameter has the same value, slight differences between the two models are observed, as e.g.~the rediffused electron component of the SEY is not included in the ECLOUD model.

\section{Numerical Investigations}
This section is divided into two parts, presenting the results of damping and collider ring, respectively.
For the damping ring simulations, we launch an
initial set of electrons with a uniform distribution 
as primary seed, and omit  
electron generations due to synchrotron radiation,
since we only wish to determine 
the onset of exponential amplification. 
By contrast, for the collider ring we consider photo-emitted
(macro-)electrons launched at the chamber wall during each bunch passage, as primary electron source.
The reason is the much lower electron-density threshold
(see Table \ref{Parameters}), where electrons from photoemission
alone could already render the beam unstable.

Simulating the collider case is more challenging also
for another reason: The transverse beam sizes in the collider arcs are much smaller than the radius of the vacuum pipe. 
Therefore, as a first step, the convergence of the simulations is confirmed by comparing the central and line electron densities for different discretization levels of the Finite Difference solver and the space charge Particle in Cell solver, 
also varying the time step and the initial macro particle size~\cite{Yaman_talk}.
We note that PyECLOUD computes the central electron density by counting the (macro)-electrons in a circle around the centre of the beam pipe, whose radius can be arbitrarily chosen. 
Therefore an additional convergence study was performed. 
The radius of the optimum computational circle was finally determined to be about $7.5$~mm.

\subsection{FCC-ee DR Injection and Extraction}
In this study, the only difference between 
DR injection and extraction is  
the different longitudinal and transverse beam sizes.
For example, the vertical beam size decreases by up to 
factor $\approx 60$ between injection and extraction. 
We consider a DR bunch spacing of 50~ns \cite{Ogur_thesis}.

\begin{figure}[!htb]
  \centering
  \subfloat[$r_0 = 10$~mm and IeD = $10^{12}$~m$^{-3}$ ]{\includegraphics[scale=0.39]{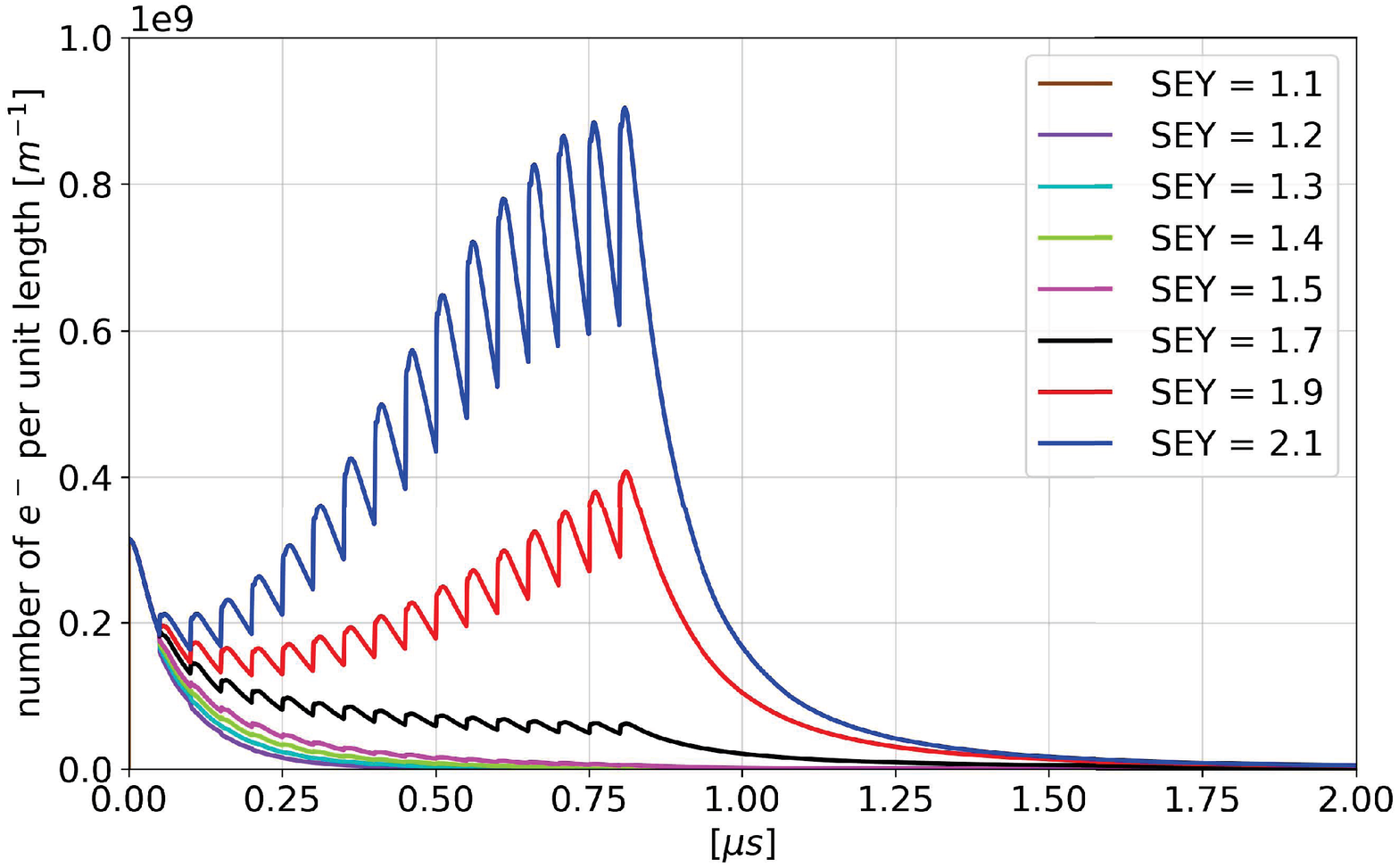} \label{fig:DR_Inj_1}} 
   \subfloat[SEY~$=2.1$]{\includegraphics[scale=0.39]{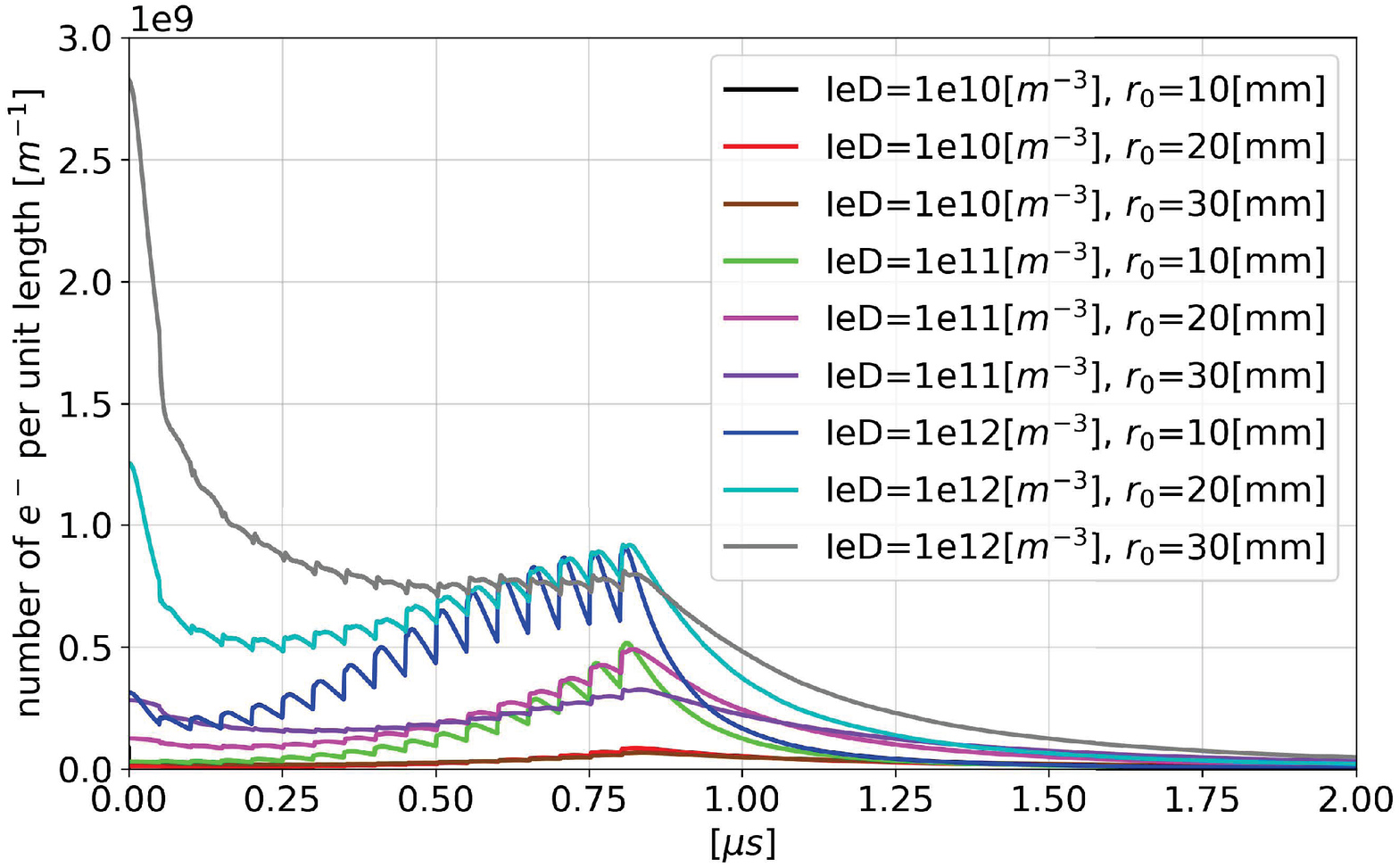} \label{fig:DR_Inj_2}}
  \caption{Simulated electron-cloud evolution from an initial uniform seed for DR injection,
  using the Furman-Pivi model. 16 bunches with 50 ns spacing are followed by a gap.}
\label{fig:FCC-ee DR_injection}
\end{figure}

In our first numerical experiment, 
we scan SEY values for the smallest radius value~($r_0$) 
of $10$~mm, and the highest initial electron density~(IeD) 
$10^{12}$~m$^{-3}$.
Figure~\ref{fig:DR_Inj_1} (left picture)
shows that for SEY~$ \leq 1.5$  
the electrons ultimately vanish 
when the computational domain is uniformly loaded at 
time step zero with cold electrons. 
In addition, no increase of the electron density from the initial seed 
was observed up to SEY$=1.7$ \cite{Yaman_121thFCCeemeeting}.
However, the SEY~$=2.1$ curve reveals a build-up-like behavior.
We investigate this case in greater detail; 
in particular, we vary the initial electron density value and the beam pipe radius, with results shown
in the right picture of Fig.~\ref{fig:DR_Inj_2}.
Here, we observe large oscillations in the electron density for the smallest beam pipe radius
and for the largest initial density. 
In this case, the electron density value reaches maximum of~$8\times 10^8$ per meter.

We note that the results for the DR at injection
were obtained with PyECLOUD and considering the Furman-Pivi SEY model, which provides larger electron build-up values than the ECLOUD model and may be considered a pessimistic scenario.
Consequently, the simulation results indicate that a serious electron build-up for the FCC-ee DR injection is not expected if the 
parameters are in the range of SEY~$\leq 1.9$, IeD $\leq 10^{12}$~m$^{-3}$ and $r_0 \geq10$~mm.

For the case of DR extraction, we only observe small changes in the number of electrons in the saturation region as compared to injection curves \cite{Yaman_et_al_FCCWeek2021}. We do not reproduce the  corresponding electron build-up curves here, as they are fairly similar to the injection case.

\begin{figure*}
    \centering
     \includegraphics*[width=0.9\textwidth]{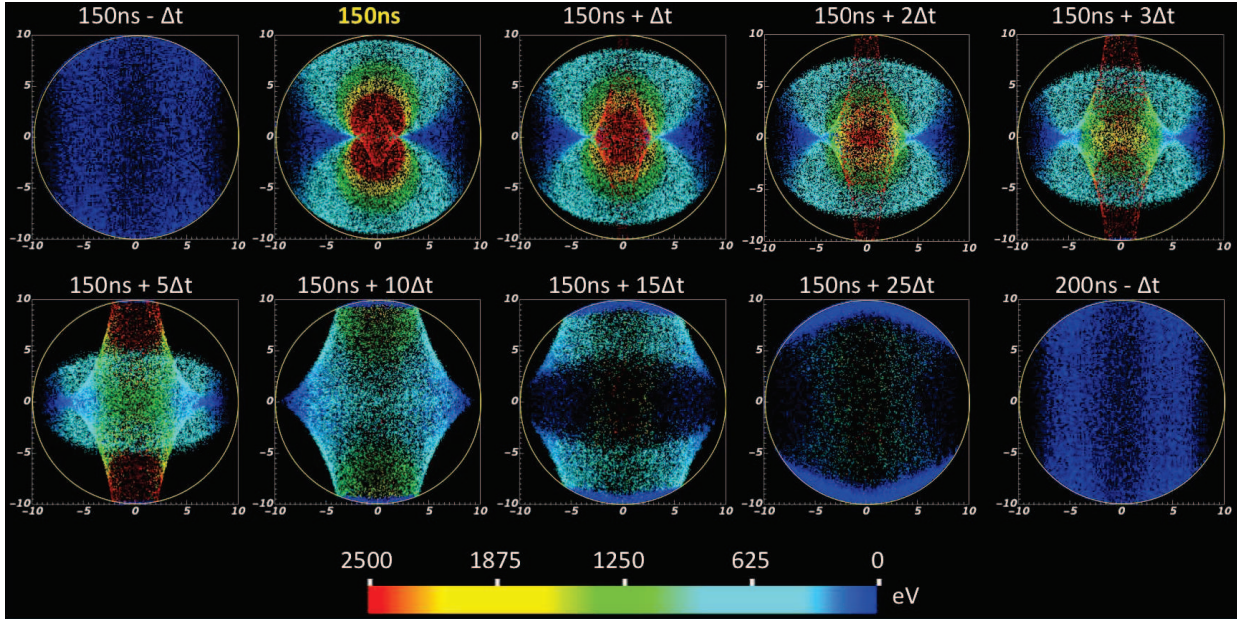}
      \caption{Kinetic energy distributions of electrons  inside the wiggler magnet chamber at particular time steps between two bunch arrivals as indicated (with $\Delta t = 62.5$~ps).  
      for FCC-ee DR extraction. }
      \label{fig:extraction}
\end{figure*}

Instead, we focus on the kinetic energy distributions of the electrons between two sequential bunch passes, specifically the time period from 150 ns to 200 ns, i.e., between the 3rd and 4th bunch passage after time zero.  
The first plot of the top row and the last plot of the second row of Fig.~\ref{fig:extraction}
show cold electrons in typical stripe formation
at the short time interval of 
$62.5$~ps before the next bunch passes. 
Energies of the electrons, which are accumulating especially around the center region of the vacuum chamber, increase up to $2.5$~keV, 
when the positively charged bunch arrives at time 
$150$~ns (second plot in top row). 
Afterwards, following the vertical field lines, 
the energized electrons reach the top and bottom sections of the chamber 
and generate new electrons though the secondary emission
process. Significant electron motion continues until most of the primary electrons have lost their energy and lower-energy secondaries emitted from the chamber wall have penetrated into the chamber.   

\subsection{FCC-ee Collider Arcs}
The electron density at the center of the beam pipe is critically important, since these electrons
can cause a vertical beam blow up, with an estimate for the threshold density value given in Eq.~(\ref{rhothr}). 
For the collider the estimated 
threshold is extraordinarily low; see Table \ref{Parameters}. 
Accordingly, in this section, simulations are mostly devoted to monitoring, and controlling, the electron density at the center of the beam pipe. 
We consider the two different SEY models discussed earlier, several possible bunch spacings, and also a few values for the bunch population, corresponding to the parameters of the Conceptual Design Report \cite{Benedikt_et_al} and to the new baseline \cite{FCCISbase}, respectively.
The photoelectron generation rate and the secondary emission yield (SEY) parameters are varied 
over realistic ranges for the FCC-ee collider arcs. 
In addition, we determine a reference electron density level for the case SEY~$\approx 0$, i.e., without any secondary emission.    
As a complementary information we also compute 
the electron line density, i.e., the electron density per unit length.

\begin{figure}[!htb]
   \centering
 	\subfloat[Minimum central electron densities for the two SEY models.] {\includegraphics*[width=0.48\columnwidth, height=0.38\columnwidth]{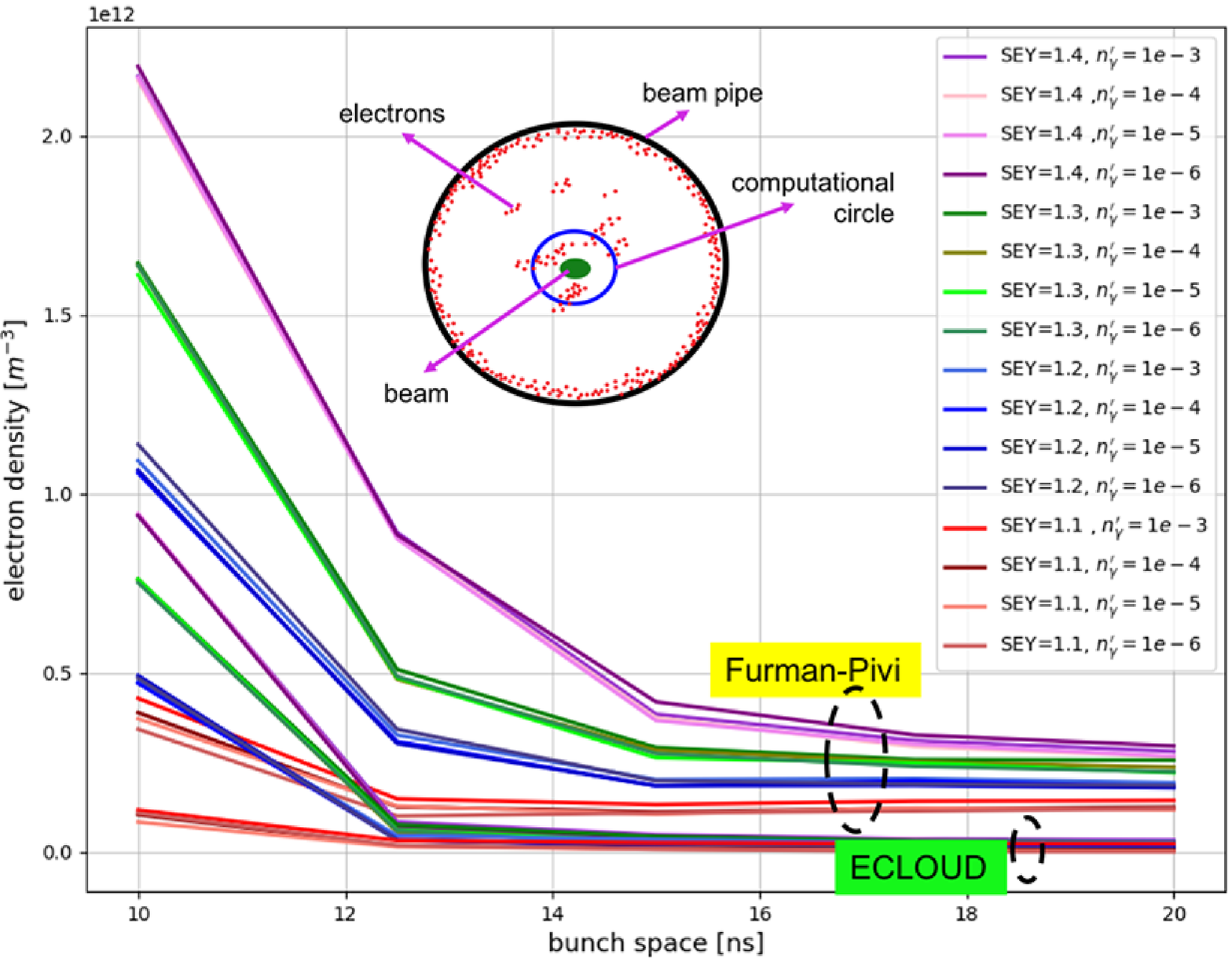} \label{fig:Min_FP_Ecloud}}
 	\subfloat[Maximum and minimum central electron-cloud density in saturation, 
 	for the Furman-Pivi SEY model.]{\includegraphics*[width=0.48\columnwidth, height=0.38\columnwidth]{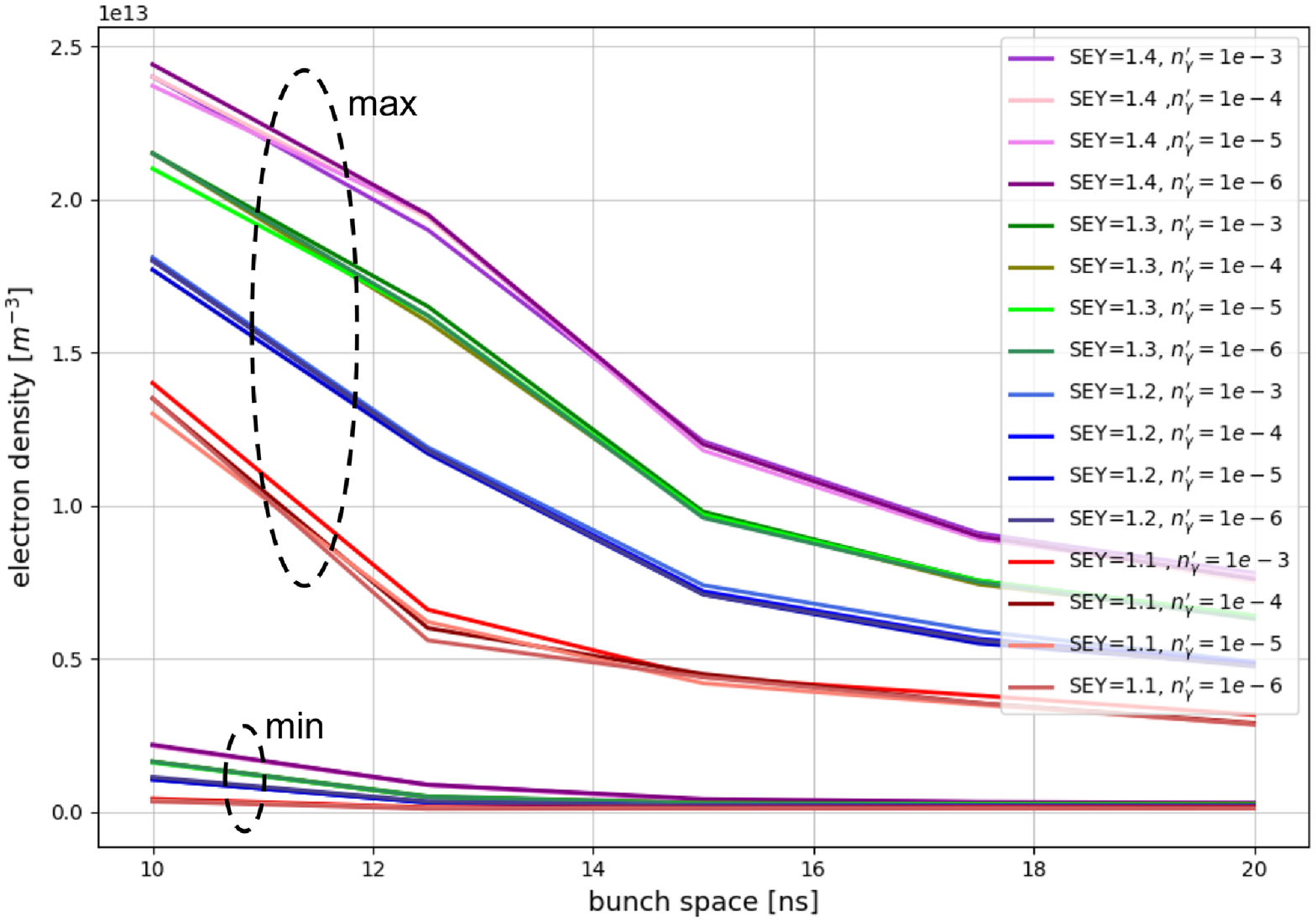}\label{fig:Max_Min_FP}}
   \caption{Electron density at the center of vacuum chamber as a function of bunch spacing.}
 \end{figure}

Our first numerical experiment for the collider arcs 
investigates the effect of choosing either the 
Furman-Pivi or the ECLOUD SEY model, 
considering various SEY 
and  $n_\gamma^\prime$ values; namely
$\mbox{SEY}=\{1.1,1.2,1.3,1.4\}$  and 
$n_\gamma^\prime = \{10^{-3}, 10^{-4}, 10^{-5}, 10^{-6} \}$~m$^{-1}$. 
Since the electron distribution is fluctuating, we calculate the average of approximate 
minimum values prior to successive bunch arrivals 
in the saturation region. With this approach, we determine the dependence of the central electron density on the bunch spacing, starting from $10$~ns up to $20$~ns. The results are displayed in  Fig.~\ref{fig:Min_FP_Ecloud}. 
In this figure, 
similarly colored curves belong to the same SEY value. The concept of the center density calculation is illustrated by the insert of the left-hand picture.  
Also in this left picture,
the curves obtained using the Furman-Pivi model exhibit 
significantly larger electron density values 
than those obtained with the ECLOUD model.
This result agrees well with the study presented in~\cite{Bellodi} 
for the sample LHC parameters. 
With either model, 
the density strongly depends on the bunch spacing.
For the Furman-Pivi model, 
at a spacing of 10 ns, the central electron cloud density 
at the moment of bunch arrival is of order $10^{12}$~m$^{-3}$, which is much higher than the threshold
density of $\sim 4\times 10^{10}$~m$^{-3}$. 
For a bunch spacing of 20 ns, the density values approach more acceptable values.
In the right picture of Fig.~\ref{fig:Max_Min_FP},
we present the mininum and maximum central density values for the Furman-Pivi model. 
For a bunch spacing of 10 ns, 
we observe a order of magnitude difference
between the minimum and maximum density.
This variation reflects the strong ``pinch'' of the electron cloud \cite{benedetto} 
with much enhanced central density near the beam during each bunch passage.
However, what matters for the instability is the 
initial (and minimum) density just prior to bunch arrival.

\begin{figure}[!htb]
  \centering
  \subfloat[$n_\gamma^\prime = 10^{-6}$~m$^{-1}$ ]{\includegraphics[scale=0.55]{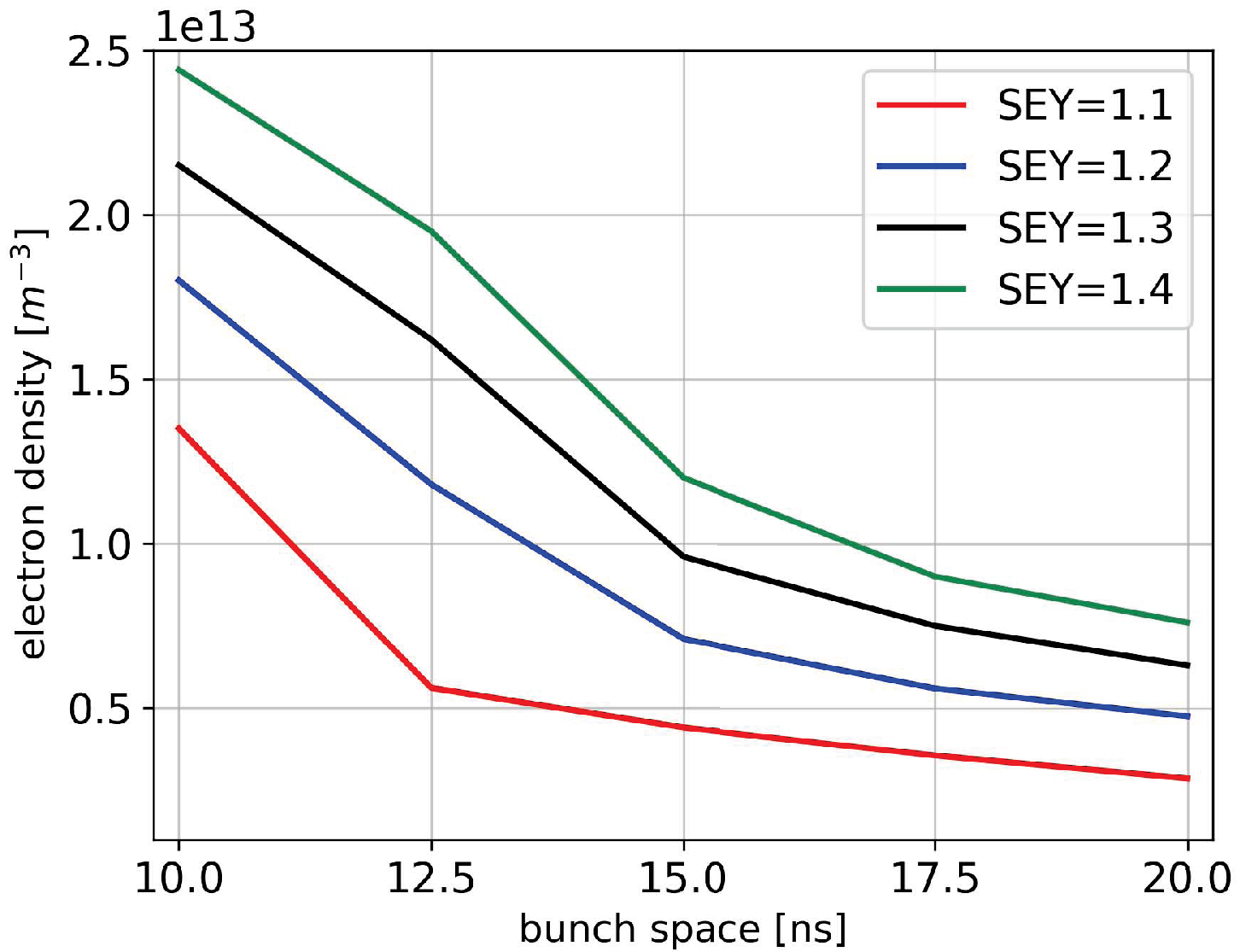} \label{var_SEY}} 
   \subfloat[SEY=1.1]{\includegraphics[scale=0.55]{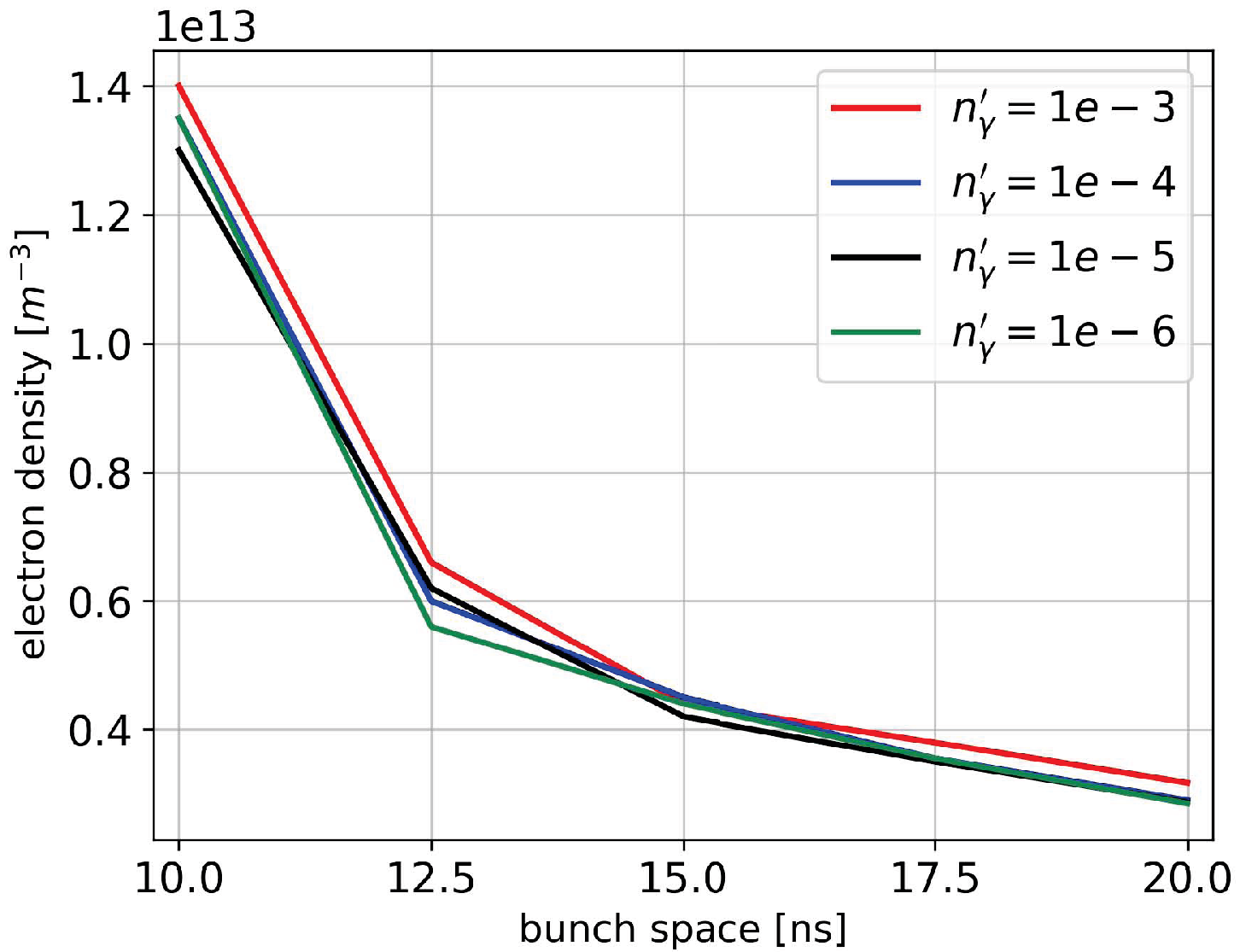} \label{var_PE}}
  \caption{Maximum central electron density for the Furman-Pivi model as a function of bunch spacing, varying either the SEY yield (left) or the photoelectron emission rate (right). }
\label{fig:various_SEY_PEY}
\end{figure}

The following simulations are performed to reveal 
whether the photoelectron generation rate 
or the secondary emission dominate the electron-cloud build up. 
We first fix the photoelectron generations with $n_\gamma^\prime = 10^{-6}$~m$^{-1}$ and 
scan the SEY value, as is shown in Fig.~\ref{var_SEY} (left picture). 
Then we hold $\mbox{SEY}=1.1$ constant, and change the value of $n_\gamma^\prime$, as is illustrated in Fig.~\ref{var_PE} (right). 
By comparing the two pictures in Fig.~\ref{fig:various_SEY_PEY}, we notice that the variation of the photoemission rate 
does not affect the center electron densities as much as 
varying the SEY value. 
However, the effect of SEY decreases for increasing 
bunch spacing.  

\begin{figure}[!htb]
   \centering
 	\subfloat[$n_\gamma^\prime = 10^{-6}$, BS=$32$~ns, $N_b=2.8 \times 10^{11}$] {\includegraphics[scale=0.55]{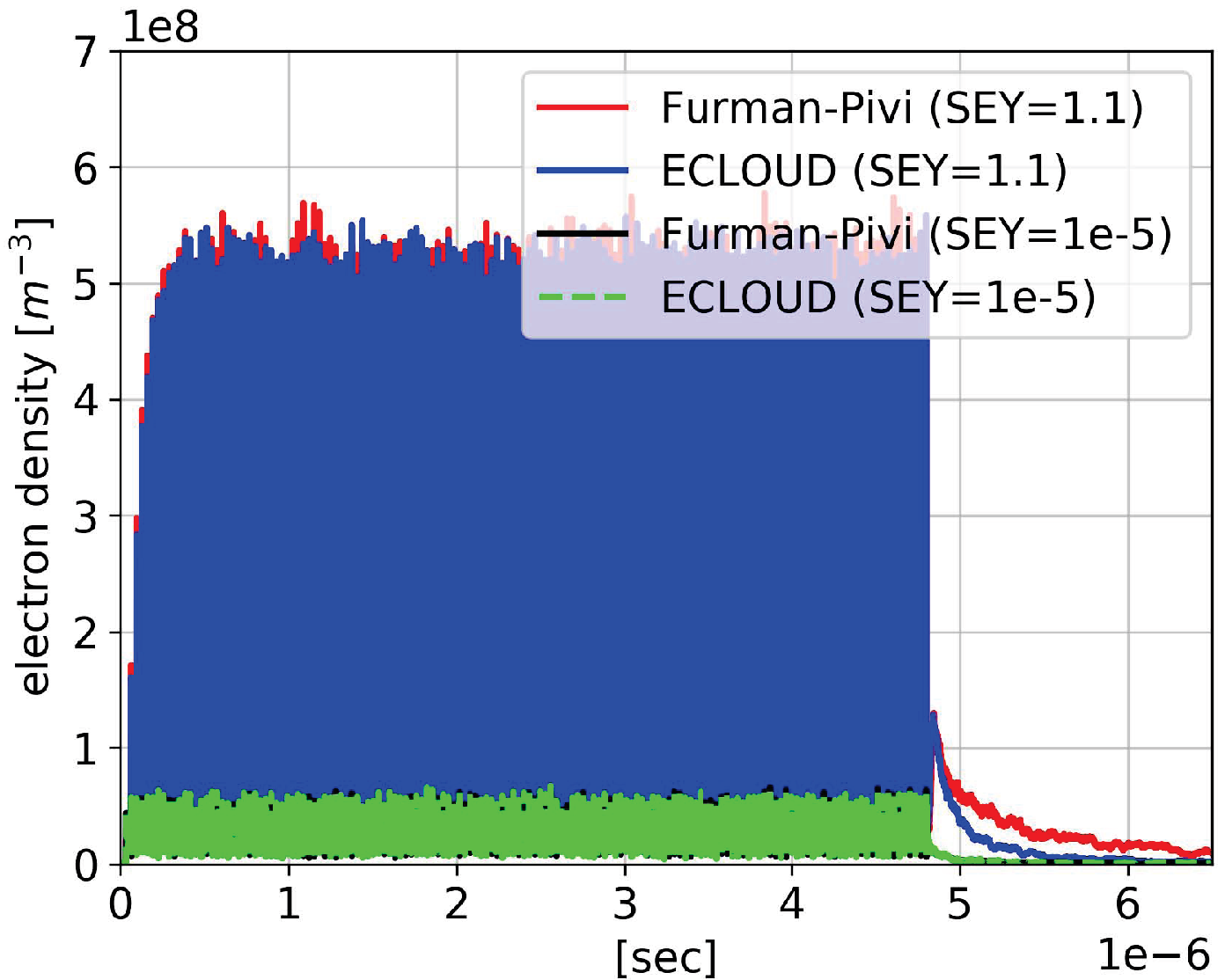} \label{fig:AgreementofSEYModels}}
 	\subfloat[Zoomed view of picture (a) at the moment of a bunch passage]{\includegraphics[scale=0.55]{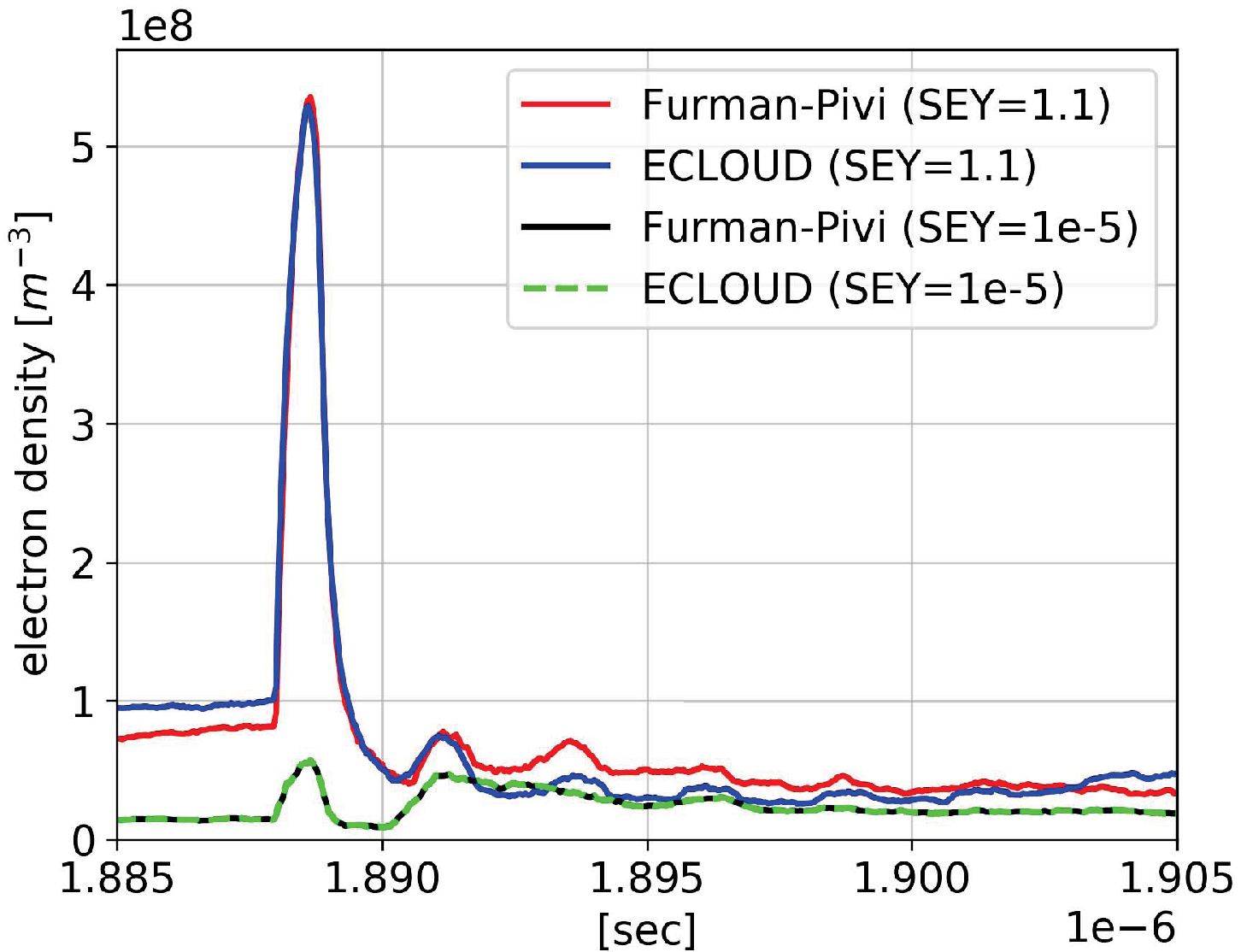}\label{fig:AgreementofSEYModelsZoom}}
   \caption{Minimum central electron cloud density versus time,  comparing SEY models and defining the reference electron formation level for $SEY\approx 0$.}
   \label{fig:coinciding}
 \end{figure}

The photoelectron generation rate becomes more prominent,
for both the Furman-Pivi 
and ECLOUD models, if we look at the minimum central density rather than the maximum, and at low values of SEY. 
Figure \ref{fig:coinciding} shows results for 
$\mbox{SEY}=1.1$ and $10^{-5}$ (essentially zero).
In this figure, BS indicates bunch spacing and $N_b$ is the bunch population. 
For SEY values larger than about 1.1,   
the influence of varying the photoelectron generation rate $n_\gamma^\prime$  from $10^{-3}$ to $10^{-6}$~m$^{-1}$  
is negligible \cite{Yaman_et_al_FCCWeek2021}.

Now we investigate the central electron density that could be reached in the ideal case of approximately zero SEY and for the lowest possible photoelectron generation rate of our parameter scan range. 
We can consider this an important reference value for the electron density. 
Figure~\ref{fig:coinciding} reveals that the reference level 
is approximately $5 \times 10^7 \, e^{-}$/m$^{3}$, and the same
for both SEY models (as the secondary emission contribution is irrelevant here). 

Next, if we choose a larger bunch spacing of $32$~ns 
and a bunch population as $2.8 \times 10^{11}$,
both values consistent with the new parameter baseline \cite{FCCISbase}, 
the central electron density obtained with 
the two SEY models agree quite 
well even for nonzero SEY yields \cite{Yaman_talk}. 
However, the overall electron density per unit length is 
not necessarily equal \cite{Yaman_talk}. 

We now examine more closely the effect of the bunch spacing and 
the number of positrons in the bunch trains 
on the electron center density level.
For this, we decrease both the bunch spacing and the bunch population to  
half their original values.
It is a common understanding that decreasing the bunch spacing 
increases the electron cloud density  
while lowering the bunch population can either 
enhance or attenuate the electron cloud build up, depending on the
initial value and other parameters such as the beam pipe radius.

\begin{figure}[htb]
  \centering
  \subfloat[SEY=1.1]{\includegraphics[scale=0.55]{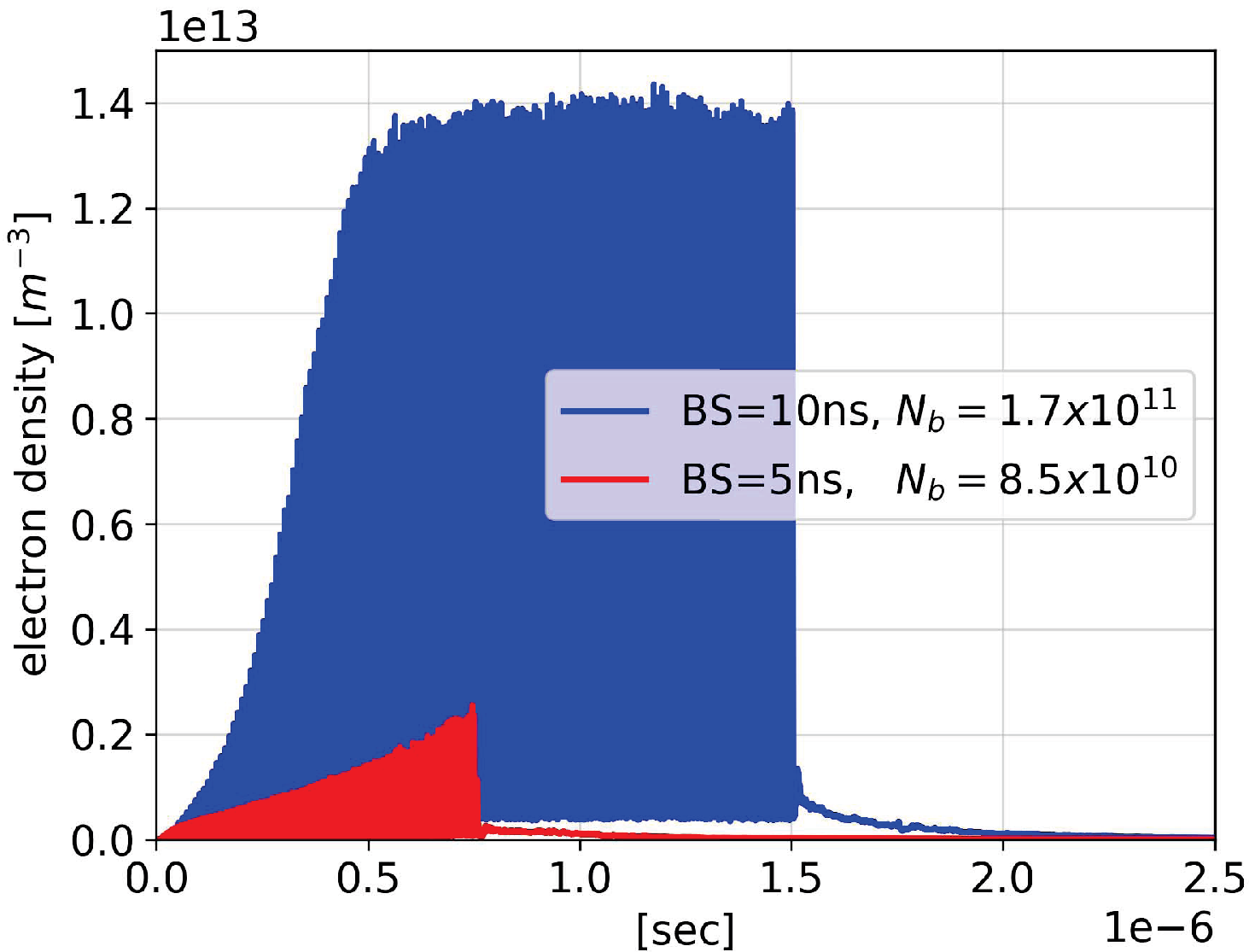} \label{PE1-3SEY11}} 
  \subfloat[SEY=1.4]{\includegraphics[scale=0.55]{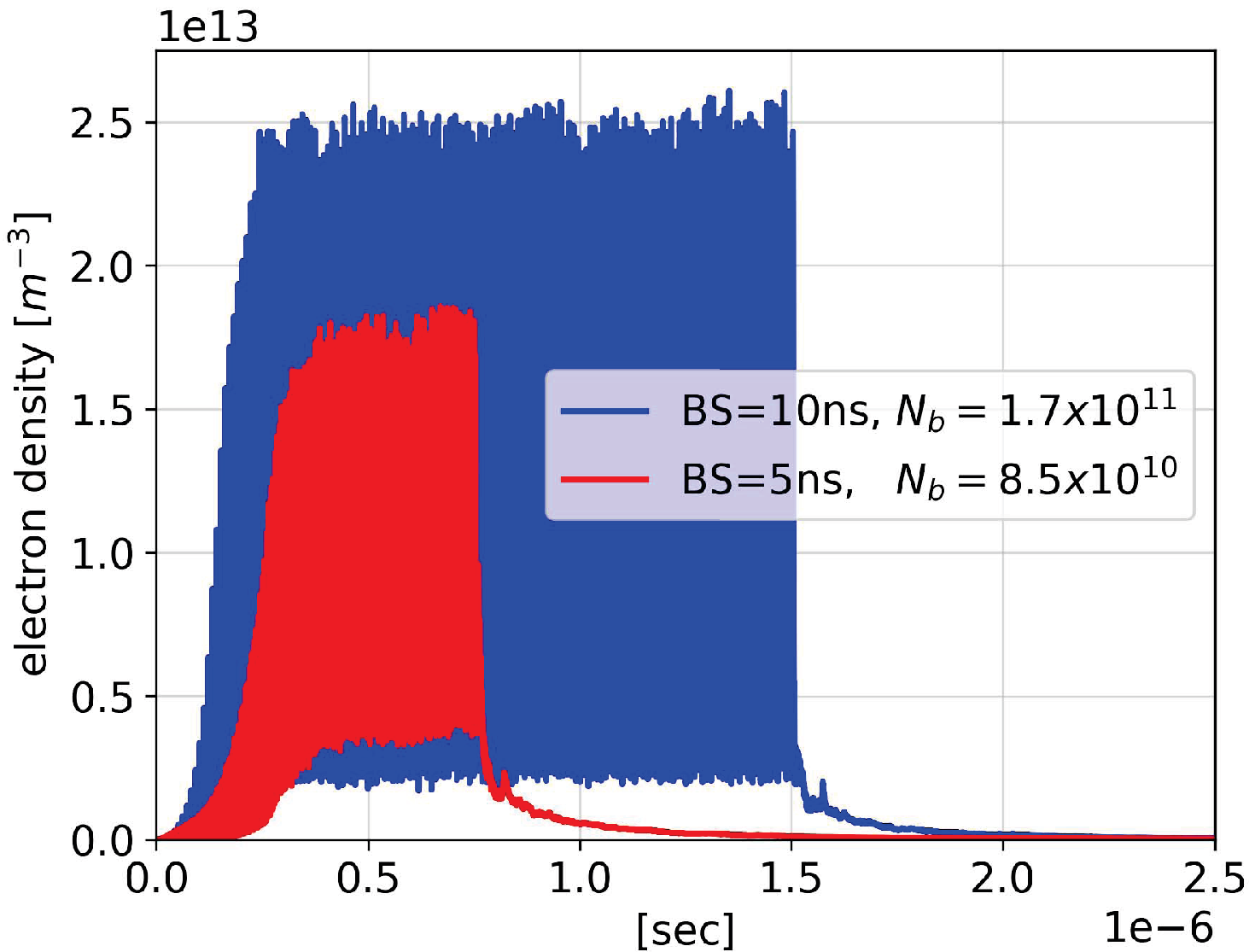} \label{PE1E-3SEY14}}
  \caption{Time evolution of the central electron density for bunch 
  spacings of 10 and 5 ns, with associated bunch population
  $N_b=1.7\times 10^{11}$~m$^-3$ and
  $8.5\times 10^{10}$~m$^{-3}$, respectively, 
  at a high electron generation rate of $n_\gamma^\prime=10^{-3}$~m$^{-1}$ (BS:bunch spacing, $N_b$: bunch population)}
\label{fig:BunchSpacingBunchPopulation}
\end{figure}

We perform the numerical experiment for the FCC-ee collider arc dipole parameters, choosing the Furman-Pivi model with SEY=1.1 and SEY=1.4. 
The simulation results are shown in  Fig.~\ref{fig:BunchSpacingBunchPopulation}.  
We can immediately conclude that the larger SEY drastically increases 
the initial speed of the 
electron-cloud build-up.
Furthermore, decreasing the bunch population and halving the bunch spacing from 10 to 5 ns has a  
beneficial effect that might permit injecting 
closely spaced bunches of lower intensity as well as avoiding an exponential electron growth in the vacuum chamber. 

Our last numerical example employs the updated FCC-ee arc dipole/drift parameters for a new 91-km layout with 4 interaction points \cite{FCCISbase}. 
In this case, the bunch population, the bunch spacing and bunch length (without beamstrahlung) are increased, compared with the FCC CDR \cite{Benedikt_et_al}, 
namely to $N_b=2.76 \times 10^{11}$,
$L_{\rm sep}=30$ or 32 ns, and $\sigma_{z}=4.32$~mm, respectively.
Additionally, also the horizontal and vertical beam sizes are increased, 
by a factor $\approx \sqrt 3$, as can be seen in Table~\ref{Parameters}.

\begin{figure}[!htb]
  \centering
  \subfloat{\includegraphics[scale=0.5]{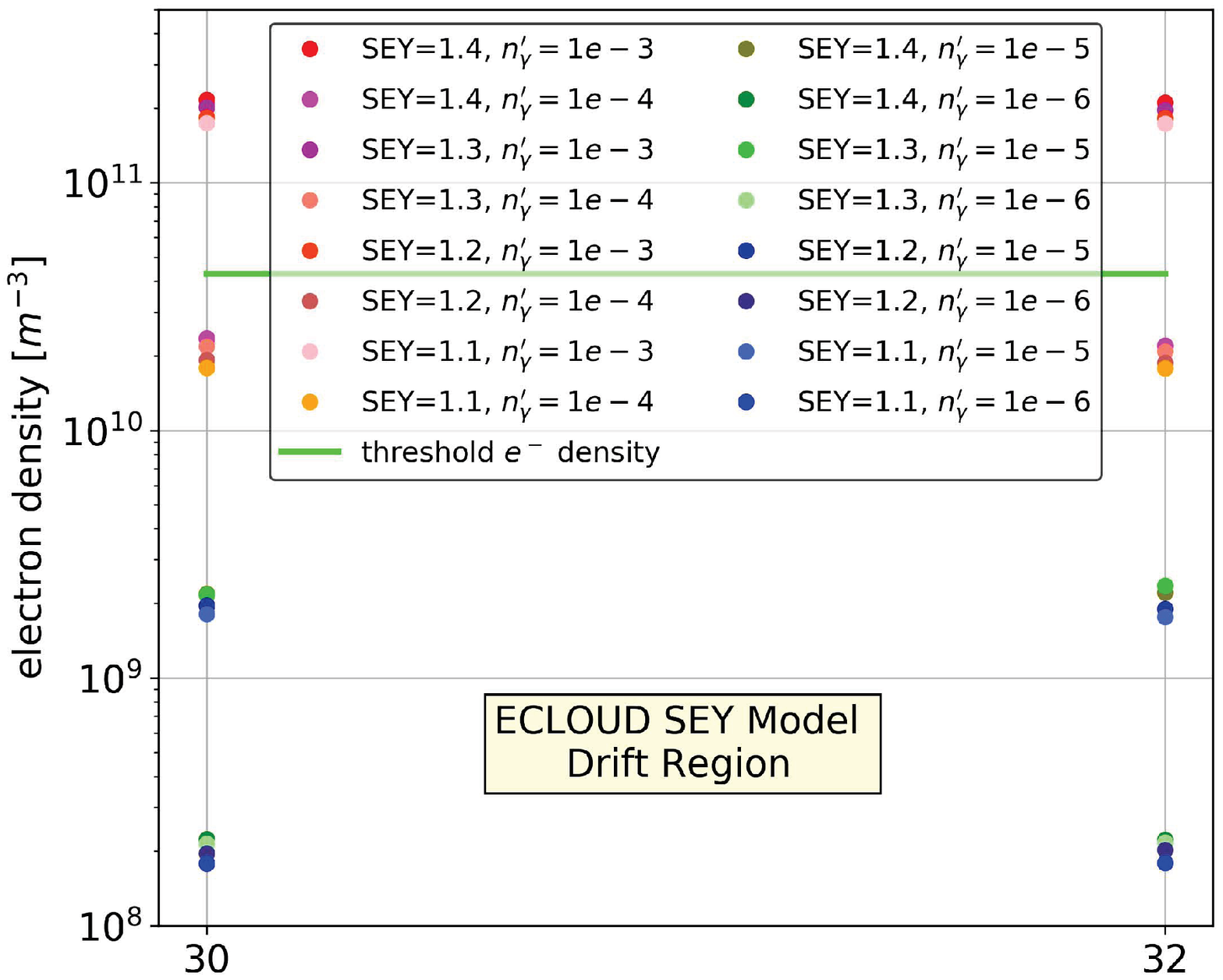} \label{4IP_EcloudDrift}} \hspace*{-0.3cm}
  \subfloat{\includegraphics[scale=0.5]{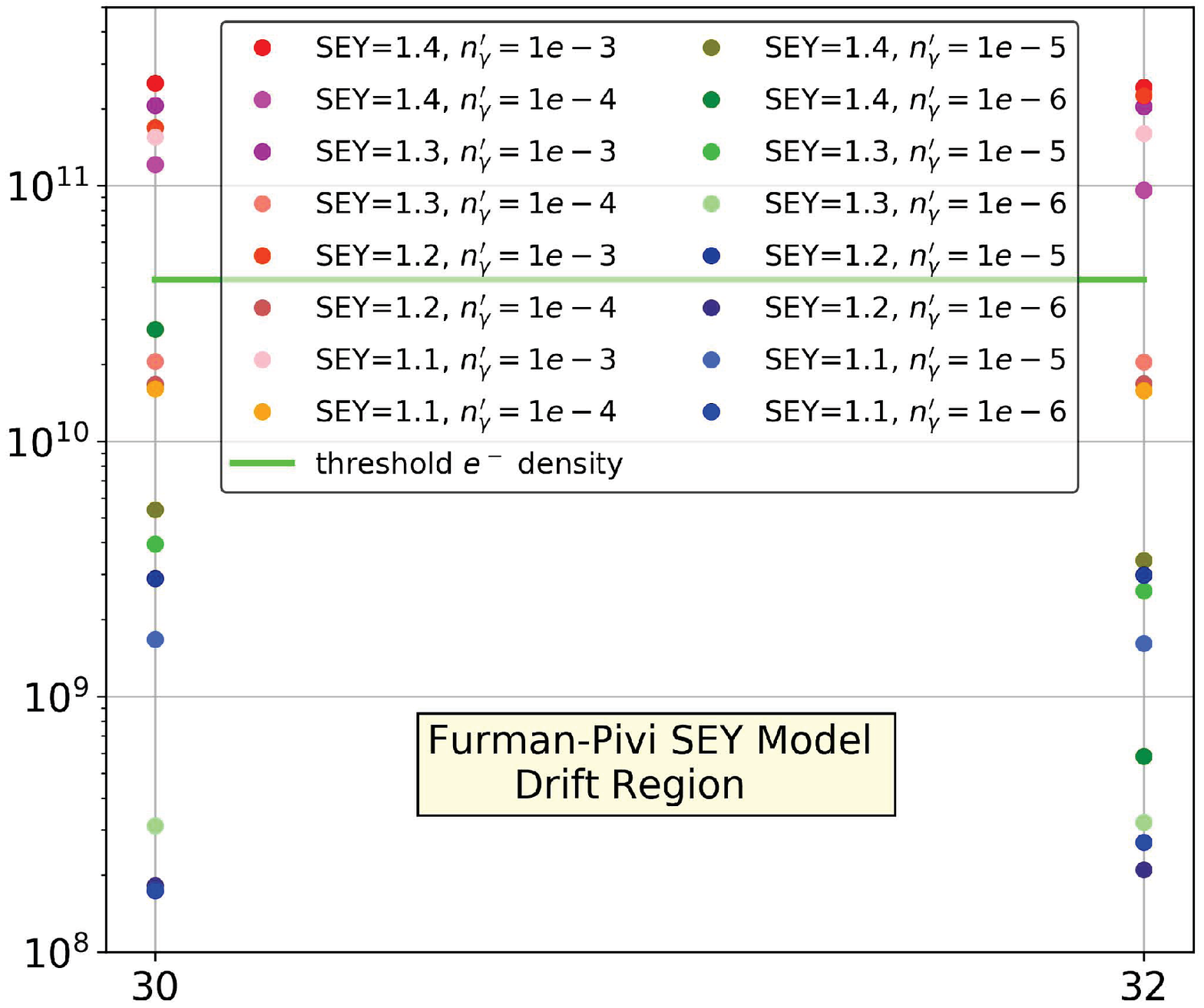} \label{4IP_FPDrift}}\\[-0.2cm]
  \subfloat{\includegraphics[scale=0.5]{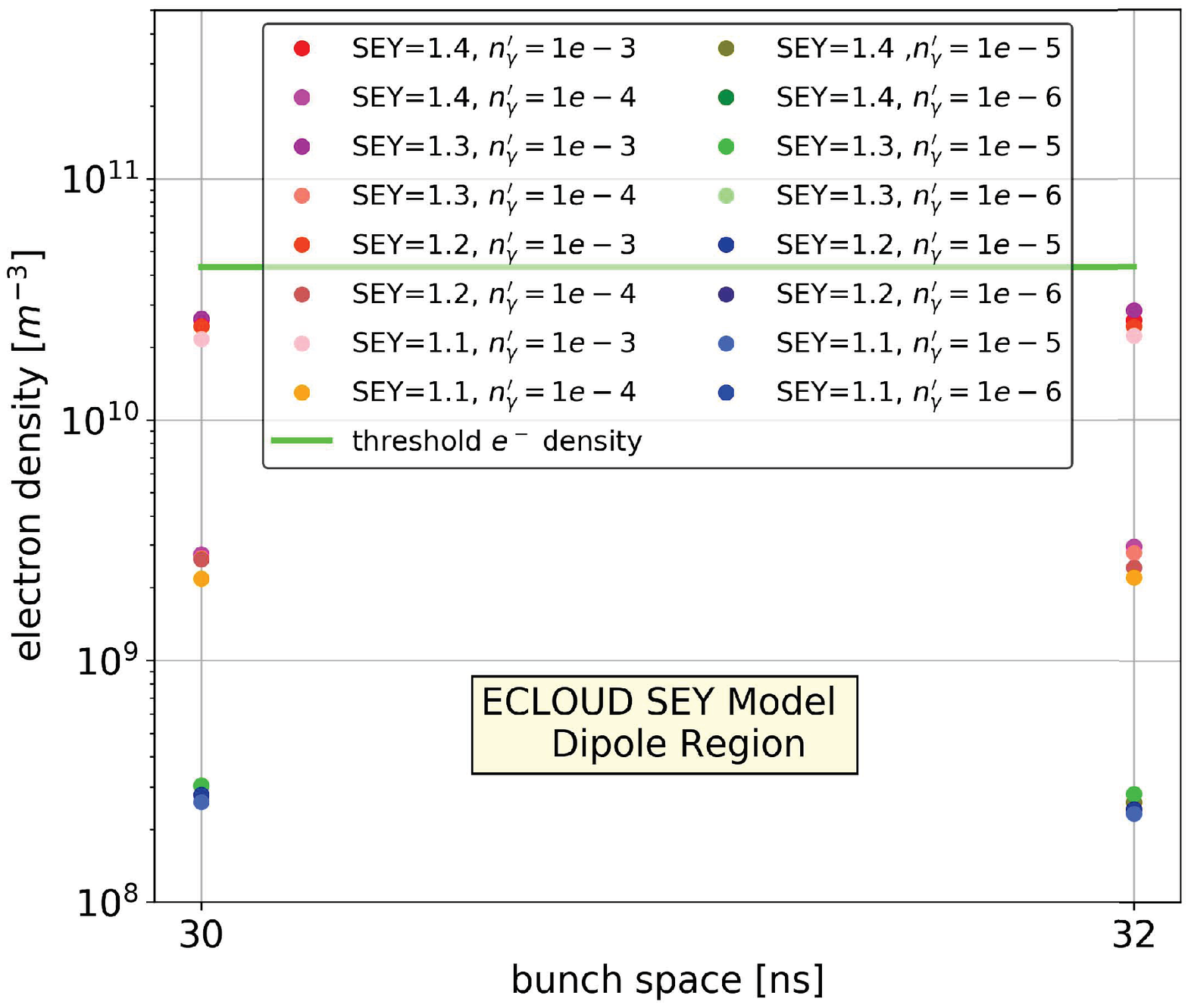} \label{4IP_EcloudDipole}} \hspace*{-0.3cm}
  \subfloat{\includegraphics[scale=0.5]{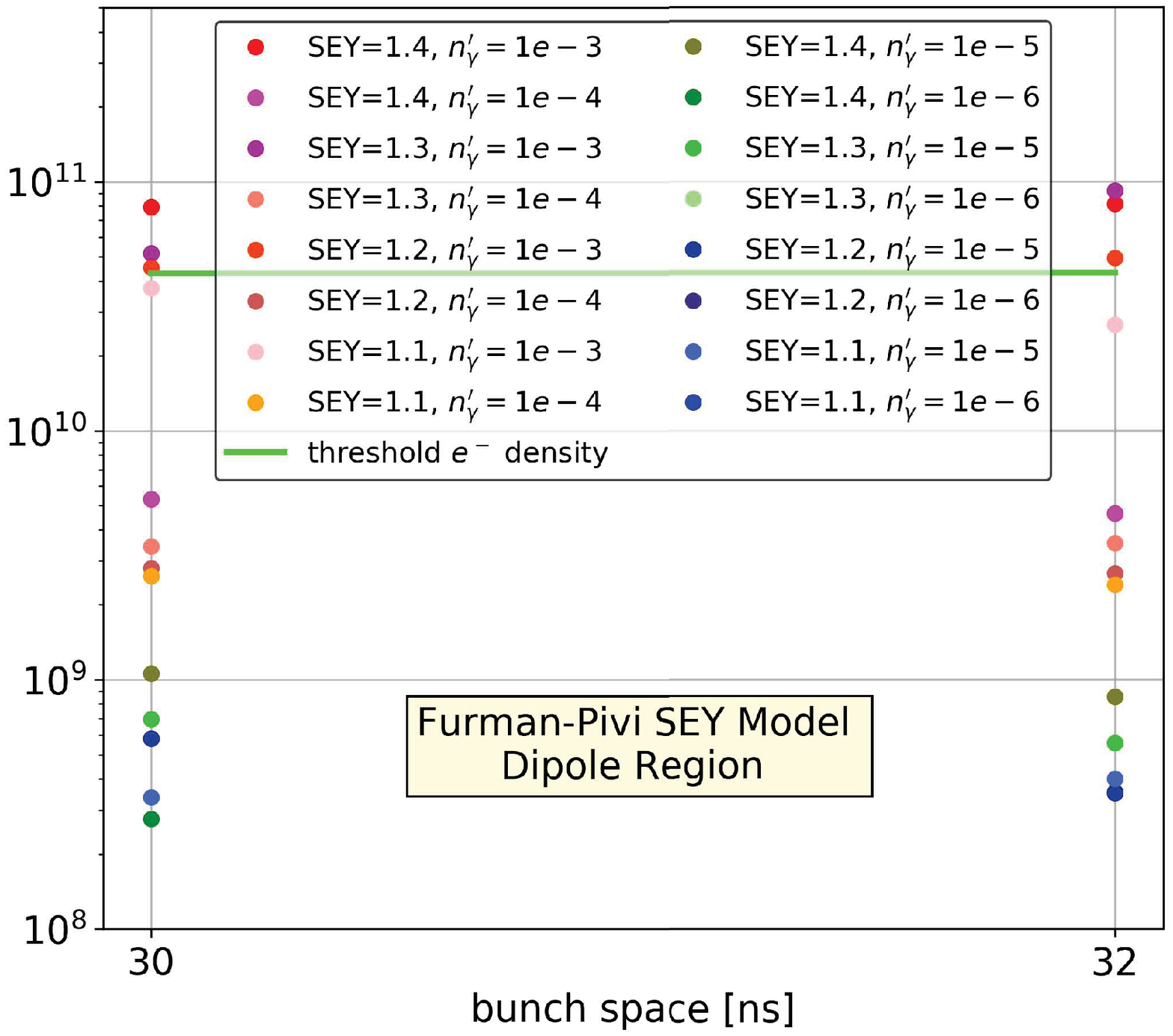} \label{4IP_FPDipole}}
  \caption{Simulated electron densities at the center of the beam pipe for the FCC-ee arc dipole \& drift regions using the new baseline beam parameters for four IPs \cite{FCCISbase}}
\label{fig:4IPResults}
\end{figure}

After substantial simulations ($\approx 450$~hours computer run time), 
we are in a position to infer certain combinations of photoelectron rate and peak secondary emission yield that result in central electron density values below the estimated threshold, 
$\rho_{\rm thr}\approx 4 \times 10^{10}$~m$^{-3}$, from (\ref{rhothr});  
also see Table \ref{Parameters}.
As a result, we find that
photoelectron generation rates $n_\gamma^\prime$ of $10^{-4}$~m$^{-1}$, $10^{-5}$~m$^{-1}$, or $10^{-6}$~m$^{-1}$ for the dipoles and $10^{-5}$~m$^{-1}$ or $10^{-6}$~m$^{-1}$ for field-free drift regions,
combined with SEY values in the range from 1.1 to 1.4, 
lead to simulated central electron densities 
lower than the estimated threshold, for both the ECLOUD and Furman-Pivi SEY 
model, and  considering either $30$~ns or $32$~ns bunch spacing, as is illustrated in Fig.~\ref{fig:4IPResults}.

\section{Conclusions}
We have reported results from electron cloud build-up simulations for the FCC-ee DR and positron collider ring.
PyECLOUD and VSim software were used to calculate the electron build up, including the electron kinetic energies.
In passing, we have also identified a few cases where 
two different SEY models lead to similar electron density values.
The photoelectron generation rate was varied over a 
realistic range of values~\cite{PrivateComm}.

For the damping ring (DR), our simulation results show that
with small beam sizes at extraction, 
the kinetic energy of electrons close to the centre of the beam pipe can exceed 2.5 keV.
Regardless, in the DR, 
both at injection and at extraction, a peak SEY value 
below 2.0 is sufficient to avoid any avalanche-like accumulation of electrons. 
Such SEY value is easily achieved with 
standard surfaces made of copper, stainless steel or NEG coating. 
Electron cloud, therefore, is not expected to be a major concern for the FCC-ee damping ring, at the present design bunch spacing of 50 ns. 
Attention would be needed only if aluminium were chosen for the DR chambers, since aluminium surfaces can exhibit SEY values well in excess of 2.0 \cite{hilleret}.   

Various combinations of SEY and photoelectron generation rate were 
considered in electron-cloud simulations for the FCC-ee collider arc dipole beam pipe, 
along with two alternative secondary emission models.
Our simulations determined the central electron density 
prior to a bunch arrival,  with bunch spacings varied between $10$ and $20$~ns for the old FCC-ee CDR parameters \cite{Benedikt_et_al}, and bunch spacings
of $30$ and $32$ ns for the new baseline \cite{FCCISbase}.

In case of the CDR parameters, the electron
density level significantly decreases with bunch spacing,.
At bunch spacings above about 15 ns it reaches acceptable 
levels of a few $10^{10}$~m$^{-3}$ with the ECLOUD SEY model, whereas
with the Furman-Pivi SEY model the electron densities 
remain about ten times higher
than the estimated instability threshold, even for a peak SEY as low as 1.1.
With the CDR parameters, 
the SEY value has a strong impact on the electron density evolution in the chamber.

However, the photoelectron generation rate gains much importance for the new parameter baseline with larger spacing and higher bunch charge.
For the updated beam parameters, a wide range of realistically 
achievable values for $n_\gamma^\prime$ and peak SEY yields
central electron density values below the estimated threshold.
For example, a photoelectron generation rate of 
$n_\gamma^\prime = 10^{-4}$~m$^{-1}$ in the dipole regions, 
along with a peak secondary emission yield of SEY=1.4,  
results in central electron density values lower than the threshold,
and this for both the ECLOUD and Furman-Pivi SEY model.
A SEY of 1.4 is well achievable, e.g.~SEY values between 1.15 and 1.35
have been demonstrated in all of the LHC arcs 
during Run 2 beam operation \cite{iadarolarun2}. 
In the FCC-ee collider, a photoelectron generation rate of
$n_\gamma^\prime = 10^{-4}$~m$^{-1}$ could be achieved,
with a chamber-wall photoemission yield of 0.1, 
if 99\% of the emitted synchrotron-radiation 
photons are effectively removed by antechambers 
and photon stops. 

Most importantly, while for the CDR parameters a model-dependent ambiguity allows no clear judgement as to whether or not an electron-cloud driven beam blow up can be avoided,  the new parameter baseline offers realistic values of photoelectron emission rate and secondary emission yield for which the electron cloud density will not
approach a critical level. 
This requirement can be taken into account in the further 
optimisation of the FCC-ee vacuum system.

\section{ACKNOWLEDGEMENTS}
The authors would like to gratefully acknowledge D.~Shatilov 
and S.A.~Veitzer for valuable discussions and the Tech-X company for providing a VSim evaluation license. This work was partially supported by the European Union's HORIZON 2020 project FCC-IS, Grant agreement n.951754.

\end{document}